\documentclass[twocolumn,showpacs,preprintnumbers,amsmath,amssymb,superscriptaddress,floatfix,nofootinbib]{revtex4}

\usepackage{graphicx}
\usepackage{epsfig}
\usepackage{epstopdf}
\usepackage{hyperref}
\usepackage{amsmath}
\usepackage{amsfonts}
\usepackage{amssymb}

\begin{document}
\title{Theoretical interpretation of the $D^+_s \to \pi^+ \pi^0 \eta$ decay and the nature of $a_0(980)$ }
\date{\today}
\author{Raquel Molina} \email{raqumoli@ucm.es}
\affiliation{Department of Physics, Guangxi Normal University, Guilin 541004, China} \affiliation{Universidad Complutense de Madrid, Facultad de F\'isica, Departamento de F\'isica Te\'orica II, Plaza Ciencias, 1, 28040 Madrid, Spain}

\author{Ju-Jun Xie} \email{xiejujun@impcas.ac.cn}
\affiliation{Institute of Modern
Physics, Chinese Academy of Sciences, Lanzhou 730000, China}
\affiliation{University of Chinese Academy of Sciences, Beijing
101408, China} \affiliation{Department of Physics and Microelectronics, Zhengzhou
University, Zhengzhou, Henan 450001, China}

\author{Wei-Hong Liang}~\email{liangwh@gxnu.edu.cn}
\affiliation{Department of Physics, Guangxi Normal University, Guilin 541004, China}

\author{Li-Sheng Geng}\email{lisheng.geng@buaa.edu.cn}
\affiliation{School of Physics and Nuclear Energy Engineering $\&$ Beijing Advanced Innovation Center for Big Data-based Precision Medicine, Beihang University, Beijing 100191, China} \affiliation{Department of Physics and Microelectronics, Zhengzhou University, Zhengzhou, Henan 450001, China}

\author{Eulogio Oset}\email{oset@ific.uv.es}
 \affiliation{Department of Physics, Guangxi Normal University, Guilin 541004, China} \affiliation{Departamento de F\'isica Te\'orica and IFIC, Centro Mixto Universidad de Valencia - CSIC, Institutos de Investigaci\'on de Paterna, Aptdo. 22085, 46071 Valencia, Spain}

\begin{abstract}

In a recent paper~\cite{Ablikim:2019pit}, the BESIII collaboration reported the so-called first observation of pure $W$-annihilation decays $D^+_s \to a^+_0(980) \pi^0$ and $D_s^+ \to a^0_0(980)\pi^+$. The measured absolute branching fractions are, however, puzzlingly larger than those of other measured pure $W$-annihilation decays by at least one order of magnitude. In addition, the relative phase between the two decay modes is found to be about 180 degrees. In this letter, we show that all these can be easily understood if the $a_0(980)$ is a dynamically generated state from $\bar{K} K$ and $\pi \eta$ interactions in coupled channels. In such a scenario, the $D^+_s$ decay proceeds via internal $W$ emission instead of $W$-annihilation, which has a larger decay rate than $W$-annihilation. The proposed decay mechanism and the molecular nature of the $a_0(980)$ also provide a natural explanation to the measured negative interference between the two decay modes.

\end{abstract}

\maketitle
\raggedbottom

In a recent BESIII experiment the $D^+_s \to \pi^+\pi^0\eta$ decay has been investigated~\cite{Ablikim:2019pit}. The dominant decay mode is found to be $D^+_s \to \rho^+ \eta$, which is a typical case of external emission ($c \to s \rho^+$) with $\bar{s}$ as spectator, and $s\bar{s} \to \eta$), and thus, is enhanced by the $N_c$ factor. In addition, two modes come from $\pi^+ \eta$ and $\pi^0 \eta$ forming the $a_0^+(980)$ and $a_0^0(980)$ resonances, respectively. These two modes are clearly seen with a cut for the invariant mass of $\pi^+ \pi^0$, $M_{\rm inv}(\pi^+\pi^0) > 1.0$ ${\rm GeV}$, which eliminates the $\rho$ contribution, and two clear peaks show up for the $a_0^0(980)$ in the $\pi^0 \eta$ invariant mass distribution and $a_0^+(980)$ in the $\pi^+ \eta$ invariant mass distribution. The combined mode has a branching ratio of about $1.5\%$ and the decay is branded as a clean example of $W$-annihilation with a rate which is one order of magnitude bigger than the typical $W$-annihilation rates.

In this work we argue that the decay mode is actually internal emission due to the nature of the $a_0(980)$ resonance as a dynamically generated state from the pseudoscalar-pseudoscalar interaction~\cite{Oller:1997ti,Oller:1997ng,Kaiser:1998fi,Locher:1997gr,Nieves:1999bx}.

\begin{figure*}[htbp]
\begin{center}
\includegraphics[scale=0.7]{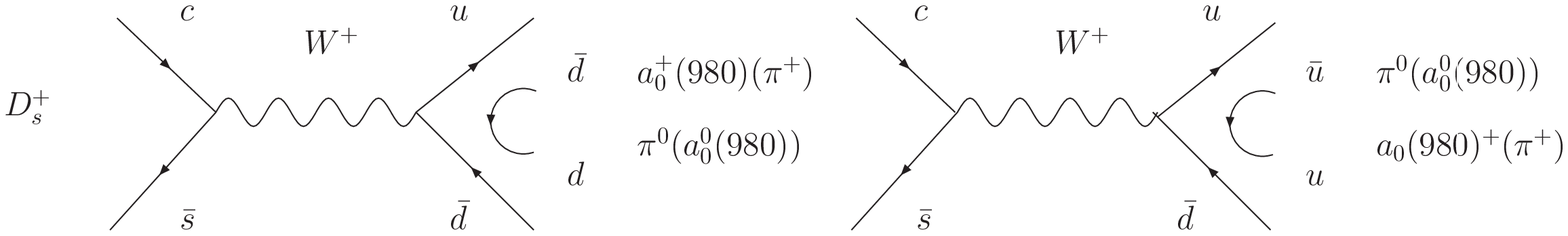}
\caption{Annihilation mechanisms assumed in Ref.~\cite{Ablikim:2019pit} for the $D^+_s \to \pi^0 a^+_0(980)$, $\pi^+ a^0_0(980)$.} \label{Fig:WAnnihilation}
\end{center}
\end{figure*}

In Ref.~\cite{Ablikim:2019pit} the $D^+_s$ decay mechanism was assumed to be given by the $W$-annihilation process depicted in Fig.~\ref{Fig:WAnnihilation}. Note that in this figure the $a_0(980)$ resonance is implicitly assumed to be a $q\bar{q}$ state. However, the advent of the chiral unitary approach to deal with the interaction of pseudoscalar mesons showed that the low-lying scalar mesons are generated from the interaction of the mesons in coupled channels and do not correspond to a $q\bar{q}$ state~\cite{Oller:1997ti,Oller:1997ng,Kaiser:1998fi,Locher:1997gr,Nieves:1999bx}. A thorough study of the $N_c$ behaviour of resonances and the properties of these scalar mesons and the ordinary vector mesons~\cite{Pelaez:2015qba} concluded that, while the vector mesons are largely $q\bar{q}$ sates, this is not the case for the low lying scalar mesons, $f_0(500)$, $f_0(980)$ and $a_0(980)$.

According to the topological classification of the weak decays in Refs.~\cite{Chau:1982da,Chau:1987tk}, the order of strength follows as $W$ external emission, $W$ internal emission, $W$-exchange, $W$-annihilation, horizontal $W$-loop and vertical $W$-loop. We first investigate whether the process can proceed via external emission. We could have the $D^+_s \to \pi^+ \bar{s} s$ external emission, which is a Cabibbo favored process, but $s\bar{s}$ has isospin zero, and upon hadronization of the $s\bar{s}$ into $K\bar{K}$, one can obtain the $f_0(980)$ final state but not directly the $a_0(980)$. The $a_0(980)$ can still be obtained through the $K\bar{K}$ propagating states due to isospin breaking because of the different $K^+$ and $K^0$ masses~\cite{Achasov:1979xc,Hanhart:2003pg,Hanhart:2007bd,Roca:2012cv}. However, this is much suppressed, and in addition there is no equivalent $\pi^0 a^+(980)$ production, while in the experiment the $\pi^+ a^0(980)$ and $\pi^0 a^+_0(980)$ modes have the same strengths~\cite{Ablikim:2019pit}.

\begin{figure*}[htbp]
\begin{center}
\includegraphics[scale=0.7]{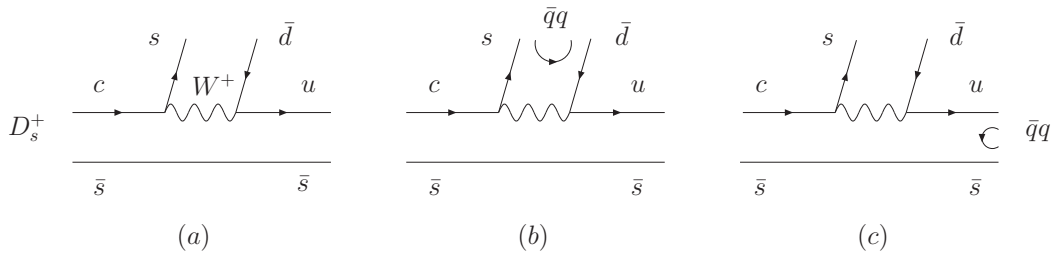}
\caption{$W$ internal emission mechanisms for the $D^+_s \to \pi^0 a^+_0(980)$, $\pi^+ a^0_0(980)$. (a) Primary step; (b) hadronization of the $s\bar{d}$ pair; (c) hadronization of the $u\bar{s}$ pair.} \label{Fig:WInternal}
\end{center}
\end{figure*}

Next we show that by means of $W$ internal emission one can obtain
the desired decay mode. The mechanism is depicted in
Fig.~\ref{Fig:WInternal}. In Fig.~\ref{Fig:WInternal} (a) we would
have two mesons in the final state. However, the $a_0(980)$ will be
produced in our picture from the interaction of two pseudoscalars.
Hence, to have $\pi a_0(980)$ in the final state we need to produce
three particles from the $D_s^+$ decay. This requires that the
$s\bar{d}$ or the $u\bar{s}$ pairs hadronize into a pair of
pseudoscalar mesons. This is done by introducing a $\bar{q}q$ state
with the quantum numbers of the vacuum. The popular way to implement
this is using the $^3P_0$ model~\cite{micu,oliver}. In the present
case, to see which pseudoscalar mesons appear from the
hadronization, the only thing we need is to impose that $\bar{q}q$
is created as a flavor scalar in SU(3). Hence, we introduce the
combination $\bar{u}u+\bar{d}d+\bar{s}s$ between the $s\bar{d}$ (or
$u\bar{s}$) quarks.

It is interesting to see which mesons appear after the hadronization in diagrams (b) and (c) of Fig.~\ref{Fig:WInternal}. For this we can write:
\begin{eqnarray}
\sum_i s\bar{q}_i q_i \bar{d} &=& \sum_i M_{3i}M_{i2} = (M^2)_{32}, \\
\sum_i u\bar{q}_i q_i \bar{s} &=& \sum_i M_{1i}M_{i3} = (M^2)_{13},
\end{eqnarray}
where $M$ is the $q\bar{q}$ matrix in $SU(3)$. If we write this matrix in terms of the pseudoscalar mesons, taking into account the $\eta$ and $\eta'$ mixing of Ref.~\cite{Bramon:1992kr} we find,
\begin{widetext}
\begin{eqnarray}
M =
\left(
\begin{array}{ccc}
\frac{1}{\sqrt{2}} \pi^0 + \frac{1}{\sqrt{3}} \eta + \frac{1}{\sqrt{6}} \eta' & \pi^+ & K^+ \\
\pi^- & - \frac{1}{\sqrt{2}} \pi^0 + \frac{1}{\sqrt{3}} \eta + \frac{1}{\sqrt{6}} \eta' & K^0 \\
K^- & \bar{K}^0 & - \frac{1}{\sqrt{3}} \eta + \frac{2}{\sqrt{6}} \eta'
\end{array}
\right),
\end{eqnarray}
\end{widetext}
and ignoring the $\eta'$ component since its large mass does not play a role in the generation of the $a_0(980)$~\cite{Oller:1997ti}, we find
\begin{eqnarray}
(M^2)_{32} &=& \pi^+ K^- - \frac{1}{\sqrt{2}} \pi^0 \bar{K}^0,  \\
(M^2)_{13} &=& \frac{1}{\sqrt{2}} \pi^0 K^+ + \pi^+ K^0 ,
\end{eqnarray}
and the three hadrons produced after hadronization of Figs.~\ref{Fig:WInternal} (b) and (c) are given by
\begin{eqnarray}
H_1 &=& (\pi^+ K^- - \frac{1}{\sqrt{2}} \pi^0 \bar{K}^0) K^+ , \label{eq:Kplus} \\
H_2 &= & (\frac{1}{\sqrt{2}} \pi^0 K^+ + \pi^+ K^0) \bar{K}^0. \label{eq:antiKzero}
\end{eqnarray}

One may wonder where is the $a_0(980)$ in these states, but this is precisely the point about dynamically generated resonances, which appear as a consequence of the final state interaction of a pair of mesons, in this case the $K\bar{K}$ component. Indeed, in Refs.~\cite{Oller:1997ti,Oller:1997ng,Kaiser:1998fi,Locher:1997gr,Nieves:1999bx} the $a_0(980)$ is generated from the $K\bar{K}$, $\pi \eta$ interaction in coupled channels in the coupled-channel chiral unitary approach ~\cite{Gasser:1983yg}. This production mechanism of scalar resonances has been utilized for explaining the $B^0_s$ decays into $J/\psi$ and $f_0(980)$~\cite{Liang:2014tia,Daub:2015xja}, $B^0$ decays into $D^0$ and $f_0(500)$, $f_0(980)$ and $a_0(980)$~\cite{Liang:2014ama}, $D^0$ decays into $K^0$ and $f_0(500)$, $f_0(980)$ and $a_0(980)$~\cite{Xie:2014tma}, among others~\cite{Oset:2016lyh,Xie:2018rqv}.

\begin{figure*}[htbp]
\begin{center}
\includegraphics[scale=0.6]{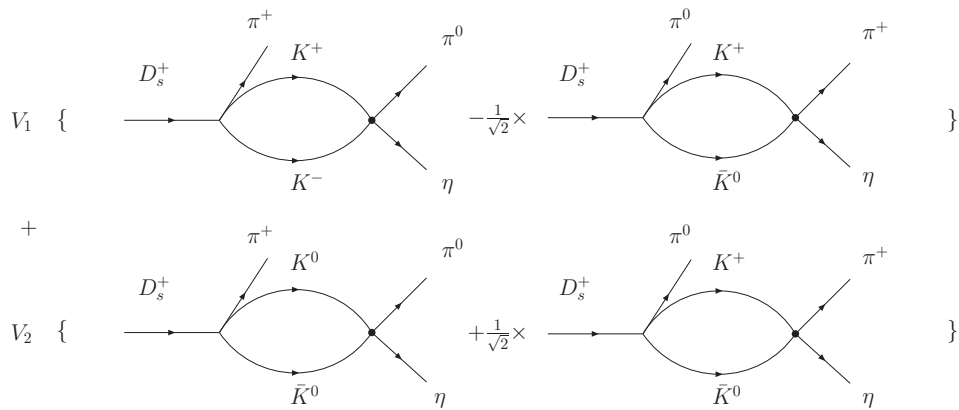}
\caption{Diagrammatic representation of the $K\bar{K}$ final state
interaction of the states $H_1$ and $H_2$ in Eqs.~\eqref{eq:Kplus}
and \eqref{eq:antiKzero} leading to $\pi^+\pi^0 \eta$ in the final
states. The $a_0(980)$ resonance appears in the scattering matrix
$K\bar{K} \to \pi \eta$ symbolized by a thick circle in the figure.}
\label{Fig:a0production}
\end{center}
\end{figure*}

The amplitudes for $\pi^+\pi^0\eta$ production through the $a_0(980)$ are given diagrammatically in Fig.~\ref{Fig:a0production}, corresponding to the two final states of Eqs.~\eqref{eq:Kplus} and \eqref{eq:antiKzero}.

Let us write the amplitude, $t$, corresponding to the mechanisms of Fig.~\ref{Fig:a0production}. For this recall the isospin multiplets that we use in our formalism, $(K^+, K^0)$ and $(\bar{K}^0, - K^-)$, and $(-\pi^+, \pi^0, \pi^-)$. Then
\begin{eqnarray}
t_{K^+K^- \to \pi^0 \eta} &=& -\frac{1}{\sqrt{2}} t^{I=1}_{K\bar{K} \to \pi \eta} , \label{tmatrixkpkm2pi0eta} \\
t_{K^0 \bar{K}^0 \to \pi^0 \eta} &=& \frac{1}{\sqrt{2}} t^{I=1}_{K\bar{K} \to \pi \eta} ,  \label{tmatrixk0k02pi0eta} \\
t_{K^+ \bar{K}^0 \to \pi^+ \eta} &=& - t^{I=1}_{K\bar{K} \to \pi \eta} , \label{tmatrixkpk02pipeta}
\end{eqnarray}
and the amplitude corresponding to Fig.~\ref{Fig:a0production} is given by
\begin{eqnarray}
t &=& V_1 [G_{K\bar{K}}(M_{\pi^0\eta}) t_{K^+ K^-
\to \pi^0 \eta}(M_{\pi^0 \eta}) \nonumber \\
&& - \frac{1}{\sqrt{2}} G_{K\bar{K}}(M_{\pi^+ \eta}) t_{K^+ \bar{K}^0
\to \pi^+ \eta}(M_{\pi^+ \eta})] \nonumber \\
&& + V_2 [G_{K\bar{K}}(M_{\pi^0\eta}) t_{K^0 \bar{K}^0
\to \pi^0 \eta}(M_{\pi^0 \eta}) \nonumber \\
&& + \frac{1}{\sqrt{2}} G_{K\bar{K}}(M_{\pi^+ \eta}) t_{K^+ \bar{K}^0
\to \pi^+ \eta}(M_{\pi^+ \eta})] \ \label{eq:decayamplitude},
\end{eqnarray}
with $M_{\pi^0 \eta}$ and $M_{\pi^+ \eta}$ the invariant mass of the $\pi^0 \eta$ and $\pi^+ \eta$ systems, respectively, where $V_1$ and $V_2$ are the weights for production of $H_1$ and $H_2$ of Eqs.~\eqref{eq:Kplus} and \eqref{eq:antiKzero}. The amplitudes in these equations above are obtained in the chiral unitary approach~\cite{Oller:1997ng}
\begin{eqnarray}
T = [1 - VG]^{-1} V ,
\end{eqnarray}
where $V$ is a $2 \times 2$ matrix with the transition potential
between the $K \bar{K}$ and $\pi \eta$ channels and $G$ (the same
$G$ function appearing in Eq.~\eqref{eq:decayamplitude}) is the loop
function of two intermediate mesons. The $V$ matrix elements are
found in Ref.~\cite{Xie:2014tma}, as well as the $G$ function for
which we use a cut off method with $q_{\rm max} = 600$ MeV to
regularize the loops~\cite{Xie:2014tma}. By means of
Eqs.~\eqref{tmatrixkpkm2pi0eta}, \eqref{tmatrixk0k02pi0eta}, and
\eqref{tmatrixkpk02pipeta}, implying isospin symmetry, we can
rewrite Eq.~\eqref{eq:decayamplitude} as
\begin{eqnarray}
&& t = \bar{V} \left[ G_{K\bar{K}}(M_{\pi^0\eta}) t^{I=1}_{K \bar{K}
\to \pi \eta}(M_{\pi^0 \eta}) \right.\nonumber \\
&& \left. - G_{K\bar{K}}(M_{\pi^+ \eta}) t^{I=1}_{K \bar{K}
\to \pi \eta}(M_{\pi^+ \eta}) \right] \ \label{eq:decayamplitudefinal},
\end{eqnarray}
with $\bar{V} = (V_2 - V_1)/\sqrt{2}$.

The finding of Eq.~\eqref{eq:decayamplitudefinal} is most welcome, because in the analysis of Ref.~\cite{Ablikim:2019pit} it is found that the two amplitudes involving the $a^0_0(980)$ and $a^+_0(980)$, leading to $\pi^0 \eta$ and $\pi^+ \eta$, appear with a relative phase close to $180$ degrees.

Actually, the former finding has its origin from a more fundamental
point. Indeed, by looking at the diagram of Fig.~\ref{Fig:WInternal}
(a) and with the isospin multiplets $(u,d)$, $(-\bar{d},\bar{u})$ we
have the states, $|s \bar{d}> = -  |1/2,1/2> $ and $|u \bar{s}> =
|1/2,1/2>$. Then in terms of total isospin we have $|s\bar{d},
u\bar{s}> = -|1,1>$. If we look now at the multiplets $\pi$ and
$a_0(980)$, we get
\begin{eqnarray}
|\pi a_0; I=1,I_3=1> = \frac{1}{\sqrt{2}} |\pi^0 a^+_0 - \pi^+ a^0_0>,
\end{eqnarray}
with $a^+_0(980)$ from $\pi^+ \eta$ and $a^0_0(980)$ from $\pi^0
\eta$, respectively. And the two states $\pi^0 a_0^+$ and $\pi^+
a^0_0$ appear with opposite sign.

Next, we show the numerical results. Since the amplitude of Eq.~\eqref{eq:decayamplitudefinal} depends on two independent invariant masses, we use the double differential width of the PDG~\cite{Tanabashi:2018oca}
\begin{eqnarray}
\frac{d^2\Gamma}{dM_{\pi^0 \eta} dM_{\pi^+ \eta}} = \frac{1}{(2\pi)^3} \frac{M_{\pi^0 \eta } M_{\pi^+ \eta}}{8M^2_{D^+_s}}  |t|^2, \label{eq:dgdm2}
\end{eqnarray}
adjusting the parameter $\bar{V}$ to the global strength of the data. By integrating Eq.~\eqref{eq:dgdm2} over each of the invariant mass variables with the limits of the Dalitz plot given in the PDG~\cite{Tanabashi:2018oca}, we obtain $d\Gamma/dM_{\pi^0 \eta}$ and $d\Gamma/dM_{\pi^+ \eta}$ and compare to the data of Ref.~\cite{Ablikim:2019pit}. The method of Refs.~\cite{Oller:1997ti,Xie:2014tma} provides good amplitudes up to 1200 MeV. The phase space requires amplitudes further away from the $a_0(980)$ resonance where they are small and play no role in the reaction. We use the same procedure as in Ref.~\cite{Debastiani:2016ayp} softening gradually the product $G_{K\bar{K}}(M_{\pi \eta})t_{K\bar{K} \to \pi \eta}(M_{\pi \eta})$ in Eq.~\eqref{eq:decayamplitudefinal} when $M_{\pi \eta} > 1.2$ GeV.

\begin{figure}[htbp]
\begin{center}
\includegraphics[scale=0.65]{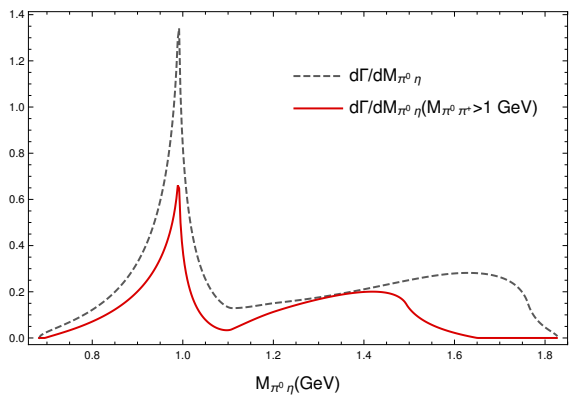}
\caption{$d\Gamma/dM_{\pi^0 \eta}$ as a function of $M_{\pi^0 \eta}$. Solid line with no $M_{\pi^+ \pi^0}$ restriction. Dashed line with the restriction of $M_{\pi^+ \pi^0} > 1$ GeV.} \label{Fig:dgdm-cut}
\end{center}
\end{figure}

The results of $d\Gamma/dM_{\pi^0 \eta}$ are shown in Fig.~\ref{Fig:dgdm-cut} (the one for $d\Gamma/dM_{\pi^+ \eta}$ is identical in the isospin limit of equal $K$ and $\pi$ masses within a given isospin multiplet). There is a neat peak, cusp like, for the $a_0(980)$ around the $K\bar{K}$ threshold, and a second broader peak at larger $\pi^0 \eta$ invariant masses. This second bump is the reflection of the $a^+_0(980)$ that the amplitude has in the $\pi^+\eta$ invariant mass from the second term in Eq.~\eqref{eq:decayamplitudefinal}.

We should recall that in the experimental analysis (see Figs. 2 (e) and (f) of Ref.~\cite{Ablikim:2019pit}) a cut $M_{\pi^+ \pi^0} > 1$ GeV is implemented to remove the $\rho^+ \eta$ contribution. Yet, this cut also removes events at higher $\pi^0 \eta$ invariant masses from the $\pi a_0(980)$ contribution. This effect is also shown in the figure.

We should note that the cusp like structure of the $a_0(980)$ is quite evident in recent high precision experiments~\cite{Rubin:2004cq,Kornicer:2016axs}, and is also remarkably visible in recent lattice ${\rm QCD}$ simulations.~\footnote{J. Dudek, talk at the Hadron 2019 Conference in Gulin, China.}

\begin{figure*}[htbp]
\begin{center}
\includegraphics[scale=0.65]{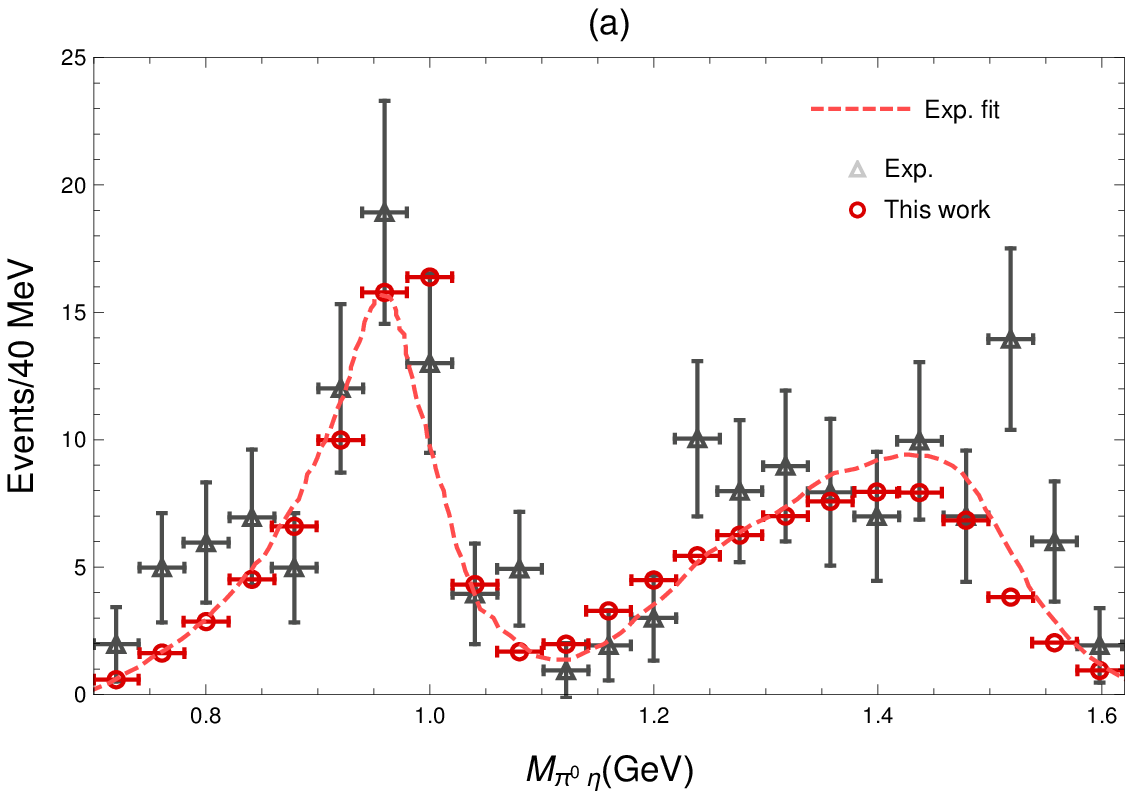} \hspace{0.5cm}
\includegraphics[scale=0.65]{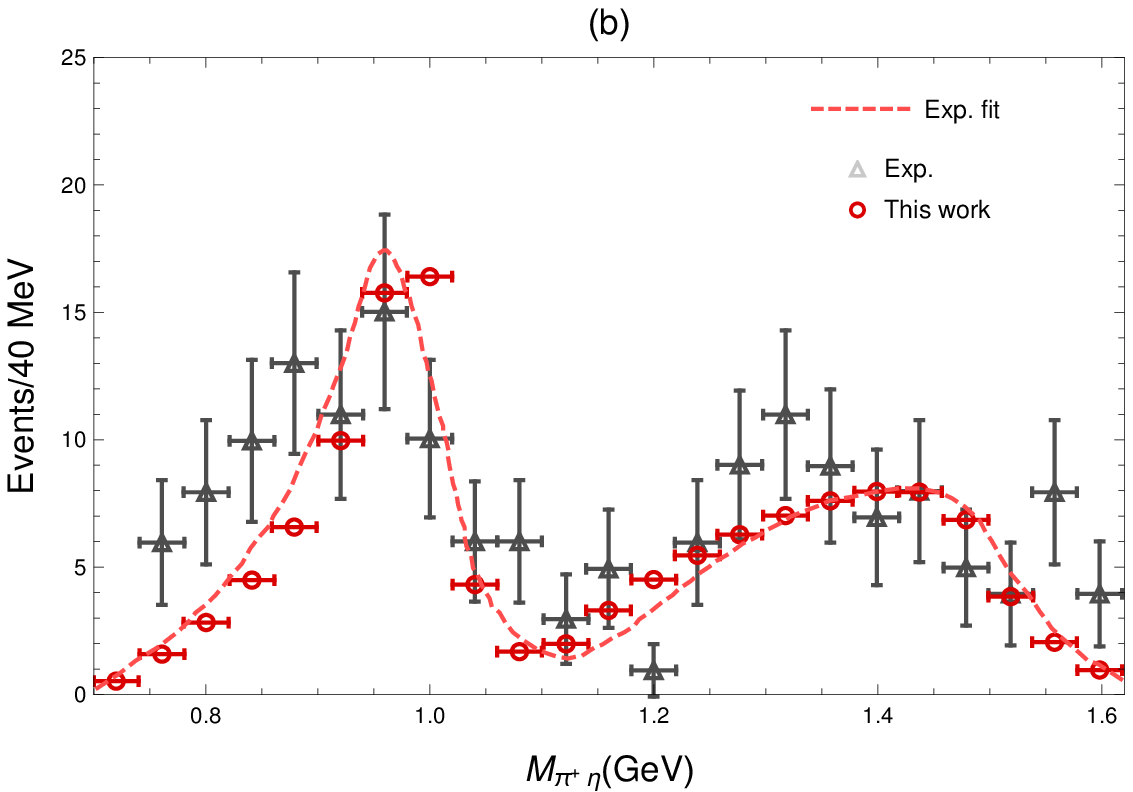}
\caption{Event distribution in $40$ MeV bins of $d\Gamma/dM_{\pi \eta}$ compared with experiment with $M_{\pi^+ \pi^0} > 1$ GeV. (a) for $\pi^0 \eta$ distribution; (b) for $\pi^+ \eta$ distribution. The dashed lines are taken from~\cite{Ablikim:2019pit} after the non $\pi a_0$ events are removed.} \label{Fig:dgdm}
\end{center}
\end{figure*}

Finally, in Fig.~\ref{Fig:dgdm} we show a distribution of events to compare with experiment. Apart from the cut $M_{\pi^+ \pi^0} > 1$ GeV we have accumulated events in bins of 40 MeV, like in the experiment integrating $d\Gamma/dM_{\pi^0 \eta}$ between $M \pm 20$ MeV, with $M$ being the centroid of the experimental bin. In this figure, $\bar{V}$ is fixed to 17.8 for comparison with experiment. The agreement with the data is fair. But, as discussed in Ref.~\cite{Ablikim:2019pit}, the data still have some contributions from other channels. Thus, the proper comparison should be done with the dashed line of the figure, which is taken from experiment after the contribution from the spurious channels is removed. The agrement with experiment is then excellent in both the $\pi^0 \eta$ and $\pi^+ \eta$ channels, which in our picture have identical strength and shape, something also found in the experiment within the experimental precision. We should comment on the small difference in the position of the $a_0(980)$ peak in the experimental analysis (dashed line in Fig.~\ref{Fig:dgdm}) and our calculation in Figs.~\ref{Fig:dgdm-cut} and \ref{Fig:dgdm}. The peak of the experimental curve comes from the parameterized $a_0(980)$ amplitude using the nominal mass of the PDG. The peak position in the theory appears at the $K \bar{K}$ threshold as a clean cusp. The position of this peak at the $K \bar{K}$ threshold and its cups-like shape are corroborated by the high statistic BESIII experiment on the $\chi_{c1} \to \eta \pi^+ \pi^-$ decay~\cite{Kornicer:2016axs}.

It is interesting to mention what happens if in Eq.~\eqref{eq:decayamplitudefinal} the minus sign is replaced by a plus sign. The distribution is drastically different, with all the strength accumulating at low invariant masses, and the $a_0(980)$ peak much less prominent. This justifies why the experimental analysis can determine this phase with high precision.

In summary, we have shown that due to the nature of the $a_0(980)$ as a dynamically generated resonance from the $K\bar{K}$ and $\pi \eta$ interaction in coupled channels, one does not need to invoke the $W$-annihilation process to explain the $D^+_s \to \pi^+ \pi^0 \eta$ through the $a_0(980)$ resonance and the process proceeds via $W$-internal emission leading to $K\bar{K}\pi$ final states, with the $K\bar{K}$ interacting to produce the $a_0(980)$ state. This mechanism solves the puzzle of the abnormally large rate observed for this decay mode compared with some genuine $W$-annihilation process like $D^+_s \to \omega \pi^+$ and $D^+_s \to \pi^0 \pi^+$. On the other hand, the good agreement with experimental data of the chiral unitary approach as shown by us here, provides extra support to the picture of the $a_0(980)$ as a dynamically generated resonance, adding to many other processes where this resonance is produced.

\section*{Acknowledgments}

R.M. and E.O. acknowledge the hospitality of Beihang University where this work was initiated. This work is partly supported by the National Natural Science Foundation of China under Grant Nos. 11735003, 11975041, 11475227, 11565007, 11847317, 11975083, 1191101015 and the Youth Innovation Promotion Association CAS (2016367). This work is also partly supported by the Spanish Ministerio de Economia y Competitividad and European FEDER funds under Contracts No. FIS2017-84038-C2-1-P B and No. FIS2017-84038-C2-2-P B, and the project Severo Ochoa of IFIC, SEV-2014-0398, and by the Talento Program of the Community of Madrid, under the project with Ref. 2018-T1/TIC-11167.

\bibliographystyle{plain}

\end{document}